\documentclass[aps,showpacs,amsmath,amssymb,prb,groupedaddress,preprint]{revtex4}

\usepackage{color}
\usepackage{graphicx,dcolumn,diagbox,bm}
\usepackage[latin9]{inputenc}
\usepackage{amsmath}
\usepackage{graphics}
\usepackage{gensymb}
\usepackage{subfigure}
\usepackage{pdftexcmds}
\usepackage{ifpdf}
\usepackage{amssymb}
\usepackage{dsfont}

\definecolor{green}{rgb}{0.15,0.85,0.35}

\allowdisplaybreaks
\makeatother
 
\begin{document}

\title{Magnetic order and exchange couplings in the frustrated diamond lattice antiferromagnet MnSc$_2$Se$_4$}
\author{K. Guratinder}\email{g.kaur@fkf.mpg.de} 
\altaffiliation[Present address: ] {Max Planck Institute for Solid State Research, 70569, Stuttgart, Germany}
\affiliation{Laboratory for Neutron Scattering and Imaging, Paul Scherrer Institute, CH-5232 Villigen, Switzerland}
\affiliation{Department of Quantum Matter Physics, University of Geneva, CH-1211 Geneva, Switzerland}
\author{V. Tsurkan} \author {L. Prodan}
\affiliation{Experimental Physics V, University of Augsburg, Germany, Institute of Applied Physics, Chisinau, Republic of Moldova}
\author{L. Keller}
\affiliation{Laboratory for Neutron Scattering and Imaging, Paul Scherrer Institute, CH-5232 Villigen, Switzerland}
\author{J.P. Embs}  \author{F. Juranyi}
\affiliation{Laboratory for Neutron Scattering and Imaging, Paul Scherrer Institute, CH-5232 Villigen, Switzerland}
\author{M. Medarde}
\affiliation{Laboratory for Multiscale Materials Experiments, Paul Scherrer Institute, CH-5232 Villigen, Switzerland}
\author{Ch. R\"{u}egg}
\affiliation{Neutrons and Muons Research Division, Paul Scherrer Institut, CH-5232 Villigen, Switzerland}
\affiliation{Department of Quantum Matter Physics, University of Geneva, CH-1211 Geneva, Switzerland}
\author{O. Zaharko}\email{oksana.zaharko@psi.ch}
\affiliation{Laboratory for Neutron Scattering and Imaging, Paul Scherrer Institute, CH-5232 Villigen, Switzerland.}

\date{\today}

\begin{abstract}
We report the magnetic properties of $A$-site spinel compound MnSc$_2$Se$_4$. The macroscopic magnetic measurements uncovers successive magnetic transitions at $T_{\rm{N1}}$= 2.04 K, followed by two further transitions at $T_{\rm{N2}}$=1.8 K and $T_{\rm{N3}}$=1.6 K.
Neutron powder diffraction reveals that both, $T_{\rm{N2}} < T < T_{\rm{N1}}$ and  $T <T_{\rm{N3}}$, orders are associated with the propagation vector $k$=(3/4 3/4 0), while the magnetic structures are collinear amplitude modulated and helical, respectively. Using neutron powder spectroscopy we demonstrated the effect of substitution of S by Se on the magnetic exchange. The energy range of the spin-wave excitations is supressed due to the chemical pressure of the $X$- ion in MnSc$_2X_4$ ($X$=S, Se) spinels.
\end{abstract}

\maketitle

\section{Introduction}{\label{Sec1}}

\begin{figure}
\centering 
\includegraphics[width=0.55\columnwidth]{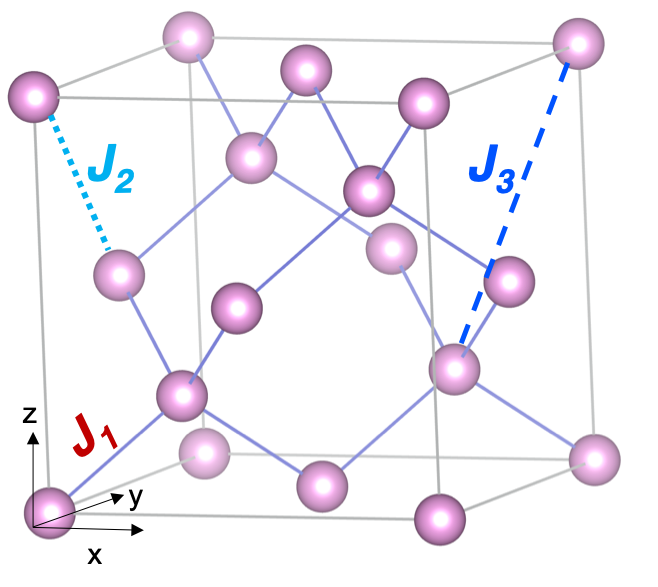} 
\caption[Diamond lattice and the exchanges]{The diamond lattice composed of $A$-sites of the spinel $AB_2$$X_4$. The  first-, second-, and third-neighbour exchange interaction terms are denoted by  $J_1$, $J_2$, and $J_3$.} \label{fig:diamond}  
\end{figure}

The $A$-sites of the $AB_2X_4$ spinel compounds constitute a bipartite diamond lattice \cite{bergman_2007, ryu_2009, schnyder_2009, wang_2015, chen_2017}, where the frustration appears due to the competition of the nearest-neighbour coupling  $J_1$ (NN), next-neighbour coupling $J_2$ (NNN) \cite{bergman_2007, chen_2017} (see Fig. \ref{fig:diamond}) or due to higher order exchange terms. In addition to the magnetic frustration, anisotropy plays a crucial role in the selection of the magnetic ground state in the $A$-site magnetic spinels\cite{lee_2008, gao2020fractional}. The well-known member of this family is MnSc$_2$S$_4$, where $S = 5/2$ Mn$^{2+}$ ions form a spiral spin liquid state\cite{gao_2017}. Recently, MnSc$_2$S$_4$ has also been found to possess a promising field-induced antiferromagnetic skyrmion lattice state \cite{gao2020fractional}. 

In an archetypal cubic spinel $A^{2+}$$B^{3+}_2$$X^{-2}_4$, the $A^{2+}$ ions occupy the tetrahedral sites and form a diamond lattice and the $B^{3+}$ ions stay in the octahedral sites forming a pyrochlore lattice. The diamond lattice of  the $A$-site spinel with magnetic ions at the (0 0 0) and (1/4 1/4 1/4) positions constitutes two interpenetrating face-centered cubic (fcc) lattices. The three most relevant exchange paths, namely $J_1$, which couples ions between two fcc lattices, $J_2$, that couples ions within same fcc lattice and $J_3$ are shown in Fig. \ref{fig:diamond}. The diamond lattice of the $A$-site magnetic spinel has four $J_1$ and twelve $J_2$ interactions.\\ The magnetic diamond lattice with only $J_1$ is not frustrated and results in a collinear antiferromagnetic order. However, the magnetic interactions become frustrated when $J_2$ is antiferromagnetic and significant, which suppresses the long-range collinear antiferromagnetic ordering and results in a spin liquid ground state. Theoretical calculations \cite{bergman_2007} showed that if the ratio of $J_2$/$J_1$ is: $0\leq J_2/J_1\leq 1/8$, the ground state will be a collinear N\'eel antiferromagnet, while for $J_2/J_1 >1/8$, it results in a degenerate spiral spin liquid ground state.\\
\textit{A}-site spinels offer multitude of exotic and novel properties to be explored experimentally as well as theoretically. Till now, $A$-site spinels unveiled a broad range of interesting materials such as cobaltates Co$_{3}$O$_{4}$ and CoRh$_{2}$O$_{4}$ \cite{suzuki_2007}; the aluminates \textit{M}Al$_{2}$O$_{4}$ with \textit{M} = Mn, Co\cite{tristan_2005, krimmel_2006, krimmel_2009, nair_2014, macdougall_2011}; and the thiospinel MnSc$_{2}$S$_{4}$\cite{fritsch_2004}. Among them, Co$_3$O$_4$, CoRh$_2$O$_4$, and MnAl$_2$O$_4$ exhibit comparatively lower frustration index, $f$= $|\Theta_{CW}|/T_N = 3.7$, 1.2, and 3.6, respectively, they lie deeply in the N\'eel phase. Recently, MnSc$_2$S$_4$ with $f$=10 realized an exotic spiral spin-liquid state \cite{radaelli_2002, bergman_2007, plumb_2016, biffin_2017}.
Extending the further insight into the intriguing physics of MnSc$_2$$X_4$ systems, here we show that MnSc$_2$Se$_4$, a sister compound, is a next candidate to host similar exotic magnetic states such as spiral spin liquid, skyrmions and vortices.\\ 
In the present work, we tune the $J_2$ and $J_1$ exchange couplings with chemical pressure by replacing S with Se. Broadly, the magnetic properties of Se- analogue resemble to MnSc$_2$S$_4$ but the exchange interaction parameters are significantly smaller. Using macroscopic magnetic measurements and neutron scattering on polycrystalline MnSc$_2$Se$_4$, an unconventional multi-step magnetic ordering has been revealed. We found a reduction of Curie-Weiss, $\Theta_{\rm CW}$ and the ordering ($T_{\rm N}$) temperatures, and reduced energy range of spin excitations on replacing S by Se. \\ 
Motivated by the findings in MnSc$_2$S$_4$ \cite{gao2020fractional} we decided to introduce the perturbations in the spin Hamiltonian and to the magnetic properties via exchange interactions. A famous example of such effects is the spinel family $AB_2X_4$, with $B$= Cr, where the size of the $A$- and $X$-site ions influences magnetic exchange interactions \cite{yaresko_2008}.  All the members of $A$Cr$_2X_4$ family reveal a clear preference to ferromagnetism on increasing the size of the $X$-site ion.\\ 
Theoretical studies of the effective exchange couplings in these spinels were primarily based on the Goodenough-Kanamori rules of the superexchange mechanism. In brief, these rules predict the sign and the strength of the interatomic spin-spin interactions controlled by superexchange mechanism. As the distance between the atoms varies, the ensuing change in exchange couplings is expected. In our case, when S is replaced by a larger Se, the lengths of both Mn-$X$ ($X$=S, Se) and Sc-$X$ bonds increase that eventually leads to change of $J_{3}$, $J_{2}$ and $J_{1}$. In the present investigation we explore a method to engineer such tuning of $J$'s in the MnSc$_2$Se$_4$ system. 
\section{Macroscopic magnetic properties}{\label{Sec2}}

The polycrystalline sample was prepared by solid-state synthesis from high-purity elements by several repeating synthesis at 1000 $^{\circ}$C. To exclude formation of oxide impurities, all technological procedures (mixing, pressing and loading of materials in ampoules) was done in an Ar box with residual concentration of O$_2$ and H$_2$O less than 1 ppm.\\
\begin{figure}  
\centering 
\includegraphics[width=0.50\columnwidth]{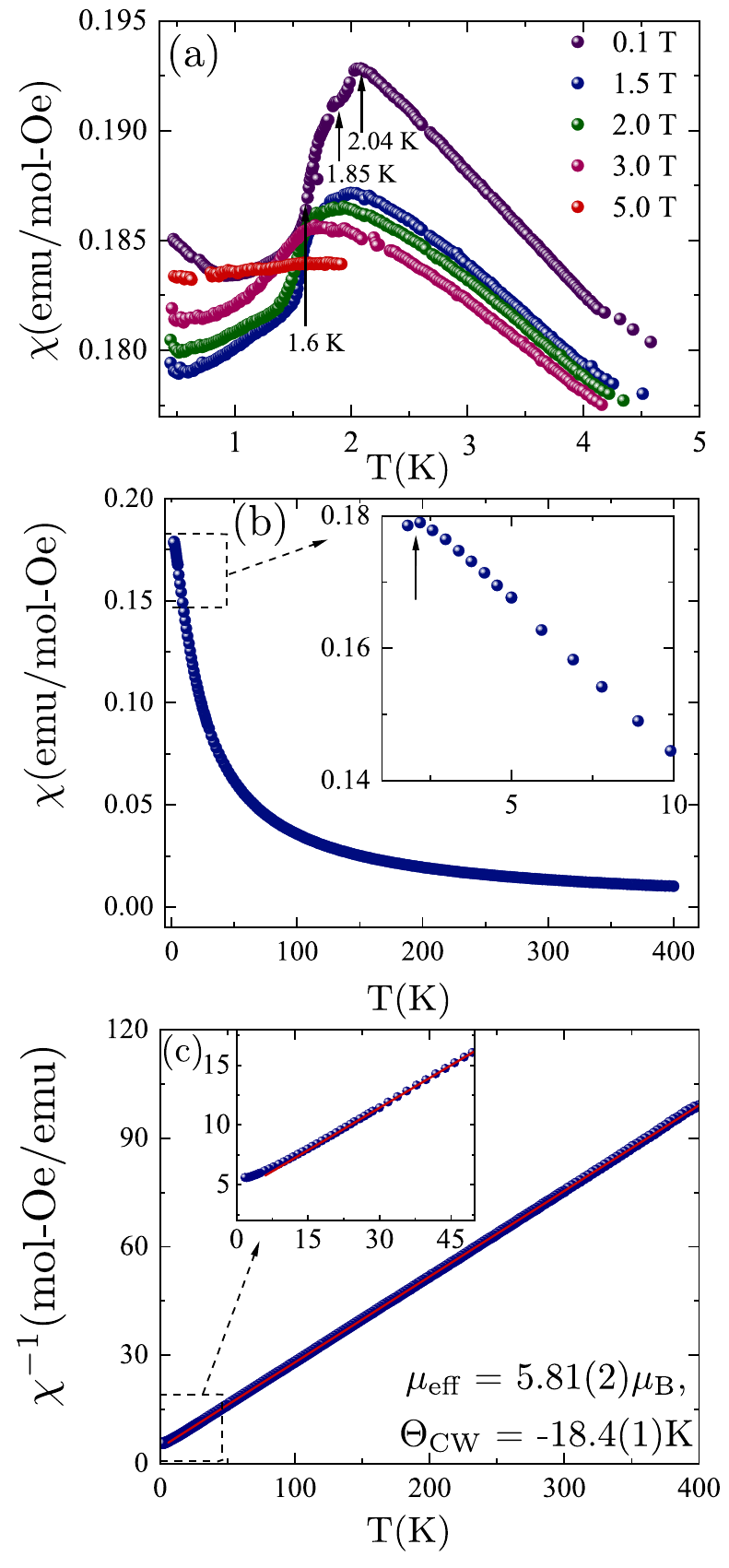}                                                                           
\caption[Macroscopic magnetization measurements for MnSc$_2$Se$_4$]{(a) Magnetization vs temperature of MnSc$_2$Se$_4$ measured at 0.1 T, 1.5 T, 2 T, 3 T and 5 T showing multiple transitions. (b) Magnetic susceptibility at H=1 T with inset indicating the transition around 2 K. (c) Inverse susceptibility with the extracted Curie-Weiss constant, $\Theta_{\rm CW}$ and an effective moment. Inset refers to the deviation from the Curie-Weiss fit close to $|\Theta_{\rm CW}|$$\sim$ 18.4 K. } 
\label{fig:squid} 
\end{figure}

The magnetization data were recorded using a commercial superconducting quantum interference device (SQUID) magnetometer (Quantum Design MPMS) with a $^{3}$He cryostat. The magnetic susceptibility data measured in the low temperature regime (450 mK $\leq T \leq$ 4.5 K) indicate three transitions at $T_{\rm N1}$ $\sim$ 2 K, $T_{\rm N2}$ $\sim$1.8 K and $T_{\rm N3}$ $\sim$1.6 K, as shown in Fig. \ref{fig:squid} (a).\\ 
These magnetic transitions get less prominent with applied magnetic field and they are not discernible anymore above 4 T. The inverse susceptibility data fitted to a Curie-Weiss law in the paramagnetic regime above 50 K, gave $\Theta_{\rm CW}$ = -18.4 (1) K. The magnetic susceptibility data measured in a broad temperature range (1.8 K $\leq T \leq$ 400 K) are shown in Fig.\ref{fig:squid} (b), Fig. \ref{fig:squid} (c) refers to the inverse susceptibility data. The effective moment $\mu_{\rm eff}$ = 5.81(2)$\mu_{\rm B}$ extracted from this fit, is close to the spin only value of 5.92 $\mu_{\rm B}$ for Mn$^{2+}$ (3$d^{5}$) $S$ = 5/2. The frustration index, $f$ = $|\Theta_{\rm CW}|/T_N$ $\sim$ 9.0, reflects the presence of significant frustration. The deviation from the Curie-Weiss behaviour below $18.4$ K (Fig. \ref{fig:squid} c (inset)) infers the presence of short-range magnetic correlations\cite{ngo_2015, pawlak_1993}. \\
\section{Magnetic correlations from neutron diffraction}
\begin{figure}
\centering 
\includegraphics[width=0.48\columnwidth]{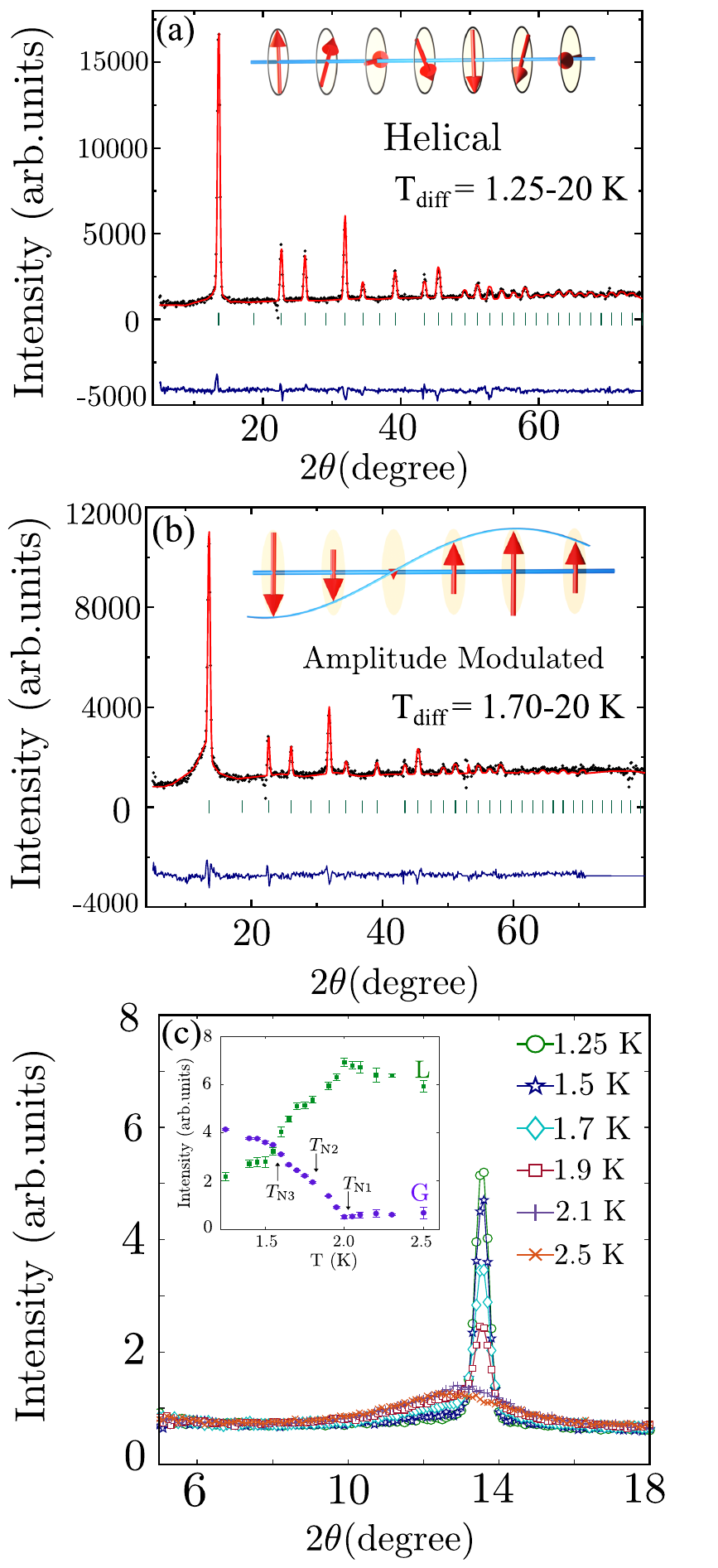}            
\caption[Comparison between helical and amplitude modulated states from DMC]{(a, b) Representative plots showing neutron powder diffraction refinement for the difference pattern obtained after subtracting the data at 20 K from low-$T$ and a constant has been added to prevent the negative intensities after the difference. Data points are shown by filled circles (black). The solid line indicates the calculated intensity (red), the vertical bars indicate the peak positions (green), the bottom line (blue) shows the difference between the observed and the calculated intensity.  (a) The refinement shown for the difference data fitted against the helical model and (b) for the amplitude modulated model. (c) Temperature evolution of the (3/4 3/4 1) magnetic Bragg peak and the (Inset) evolution of the Lorentzian (L) and the Gaussian (G) peak components.}
\label{fig:refinement} 
\end{figure}

{\label{Sec4}}
Neutron powder diffraction patterns were collected on DMC, SINQ at PSI with an incident neutron wavelength of 2.46 \AA. The experiment was performed with $\sim$ 4.58 g of polycrystalline sample, which was inserted into an air-sealed vanadium  sample can (V-can) with an inner diameter of 8 mm. 
To obtain the desired low temperatures up to 1.3 K, we have used a He-cryostat with a roots pump. The patterns have been recorded at several temperatures that were chosen based on the results from the macroscopic magnetization measurements. \\ 
The obtained diffraction data have been analysed using the Fullprof program \cite{fullprof1993}. Firstly, the nuclear structure was refined against the data taken at 20 K, which confirms the space group $Fd\bar{3}m$ (227) and lattice constants: $a=b=c=$11.0873(6) \AA. The Fig. \ref{fig:refinement} (a, b) shows the difference pattern obtained by subtracting the data taken at  20 K (uncorrelated state) from the data at low temperatures (correlated state), here, the low temperatures refer to 1.25 K and 1.7 K.\\
We refined the difference data with the propagation vector $k$=(3/4 3/4 0) using the helical and the amplitude modulated models developed for MnSc$_2$S$_4$ \cite{gao_2017}. The different quantities, commonly termed as agreement factors are calculated to identify the quality of the fit. \\
From our refinement of the difference data, we obtained, for the helical model magnetic Bragg factor ($R_{B}$)=10.3 \%, profile factor ($R_{p}$)=5.9 \%, weighted profile factor ($R_{wp}$)=6.8\% and the goodness of fit, $\chi^{2}$ = 2.6. This model fits very well near to $T$=1.25 K but fails to converge for higher temperatures. Then the amplitude modulated model is used to describe the data and we conclude that this model is fitting well at and above $T$=1.7 K. The agreement factors are $R_{B}$ = 18\%, $R_{p}$ = 5.67 \%, $R_{wp}$ =6.83 \% and $\chi^{2}$ = 2.82. In the helical structure the moments attain an equal value of 3.57(5) $\mu_{\rm B}$ and rotate normal to the propagation vector $k$=(3/4 3/4 0).
In the amplitude modulated structure the maximal moment value is 2.94(2) $\mu_{\rm B}$ and all moments point normal to $k$, along the (1-10) axis.\\
We analyse now the diffuse scattering present in the patterns. The strongest magnetic intensity of MnSc$_2$Se$_4$ associated to the first magnetic reflection ($\pm$3/4 $\pm$3/4 \ 0) has two components. These two components are, a  resolution limited Gaussian and a broad diffusive Lorentzian features. The temperature dependence of these components is also presented in Fig. \ref{fig:refinement} (c). The two features reveal the long-range (LRO) and short-range (SRO) order in the sample, respectively. The narrow Gaussian Bragg peak dominates at low temperatures but then vanishes around 2 K, which is consistent with the low-$T$ anomaly of the magnetic susceptibility (Fig. \ref{fig:squid}). It is a clear indicator of the LRO in MnSc$_2$Se$_4$ present below $T_{\rm N1}$= 2.04 K. On the other hand, the Lorentzian contribution extends to higher temperatures and can be ascertain up to 18.4 K corresponding to $|\Theta_{ \rm CW}|$. Hence, these results substantiate that short range correlations evolve below $\Theta_{\rm CW}$ and even tend to persist around $T$=1.6 K but due to the frustration effects, the LRO is restrained down to $T_{\rm N1}$= $\Theta_{\rm CW}$/10.
\section{Dynamic magnetic correlations}{\label{Sec5}}
\begin{figure}       
\includegraphics[width=0.65\columnwidth]{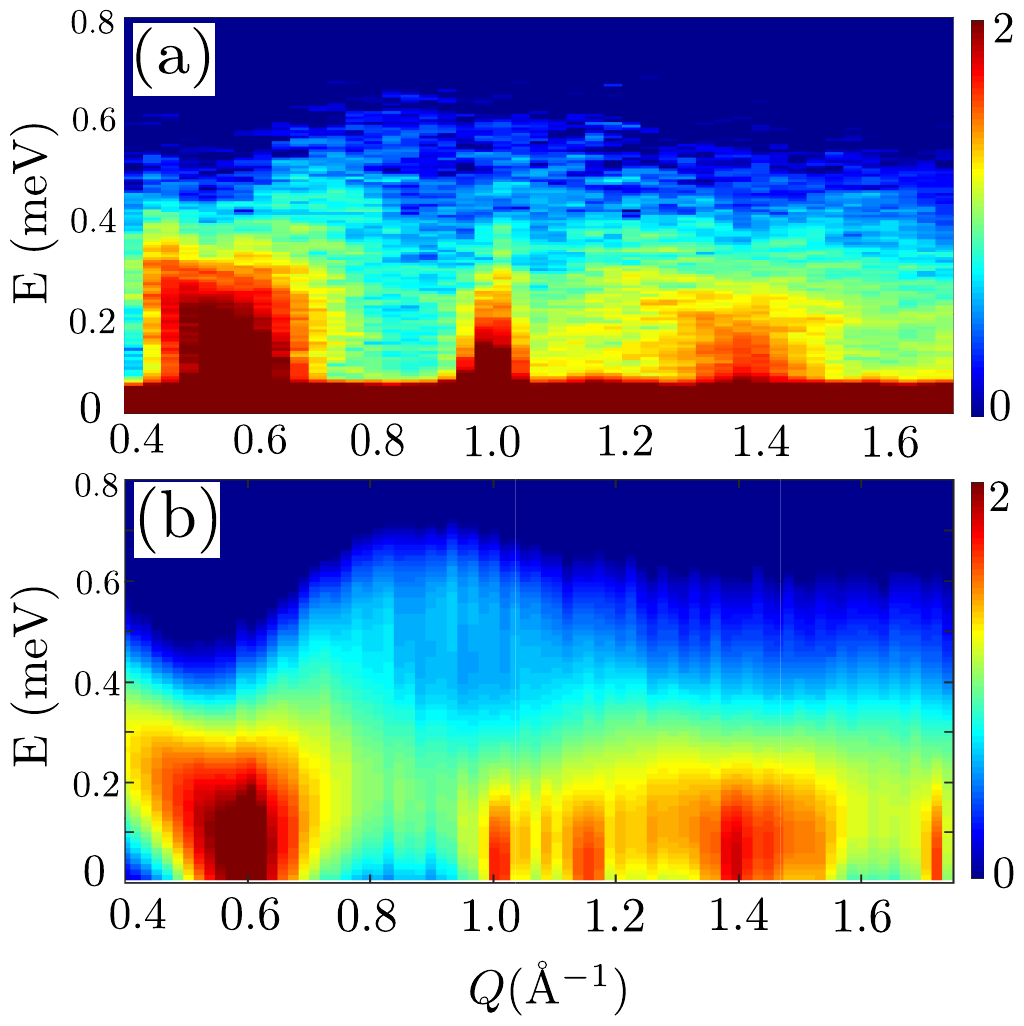}  
\includegraphics[width=0.9\columnwidth]{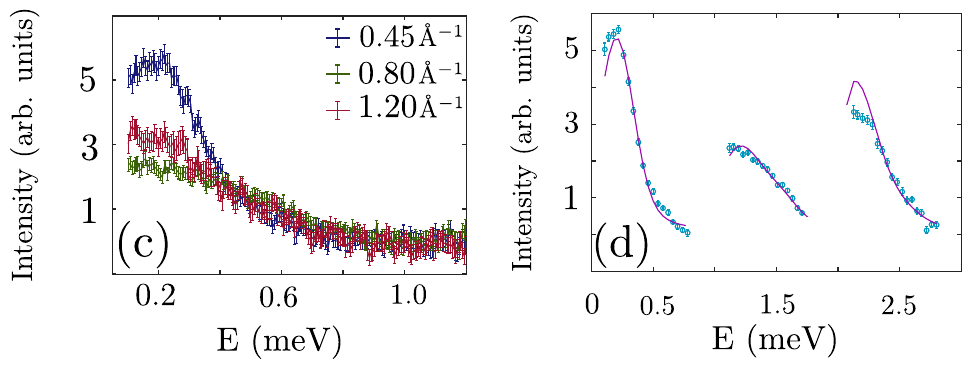}                            
\caption[Comparison between the experimental and theory results from FOCUS]{(a) Experimental excitation spectrum measured on FOCUS with $E_i$=2.47\,meV at T=1.6 K and (b) Fitted results with SpinW with $Q$= 0.45, 0.8, and 1.2 \AA$^{-1}$. (c) For the refinement, three different $Q$-strips chosen with $\delta Q$=0.05 \AA$^{-1}$ and (d) SpinW fit for the three $Q$ strips. Data points are shown by blue circles. The solid violet line indicates the calculated intensity.} 
\label{fig:focus_results} 
\end{figure}
Inelastic neutron scattering data were collected with the incident neutron wavelength of 5.75 \AA  (E$_i$=2.47\,meV) on the time-of-flight spectrometer FOCUS, SINQ at PSI. For this experiment $\sim$ 4.2 g powder sample was filled into  an Al-can of an inner diameter of 8 mm which was then cooled to T=1.6 K, the base temperature of a He-cryostat. For the data reduction, we employed the standard DAVE program \cite{DAVE}, which converts the TOF data into energy transfers at constant $Q$. To account for the background, data for an empty Al-can were subtracted. We observed spin wave excitations at the base temperature of 1.6 K, emerging from the Bragg positions and reaching the energy maximum of  0.7 meV at $Q$=0.6 \AA$^{-1}$, see Fig. \ref{fig:focus_results} (a).\\
To understand these excitations, we performed calculations based on the linear spin wave theory using the SpinW package\cite{toth_2015}. We used the $J_{1}-J_{2}-J_{3}$ model to fit the excitations at $Q$= 0.45, 0.8, and 1.2 \AA$^{-1}$ as shown in the Fig. \ref{fig:focus_results} (b-d). The parameters used for the fitting are the exchange parameters, peak width and background. The best fitted parameters are found to be $J_1$= -0.29 K, $J_2$= 0.36 K and $J_3$= 0.047 K.\\
The ratio of $|J_2$/$J_1|$ for MnSc$_2$Se$_4$ lies in the regime of the highly frustrated system as the threshold to realize the spiral spin-liquid state is given by $J_2$/$J_1 >1/8$\cite{bergman_2007}. The value of the exchange parameters, $J_2$ and $J_3$ are significantly smaller and $J_1$ is nearly similar, relative to the sulphur analogue\cite{gao2020fractional}. These results can be understood intuitively by addressing the effect of the larger size of Se$^{2-}$ ion and, as a consequence, the strength of exchange interactions weakens. The fitting results reflect that the substitution of the $X$-ion (S $\rightarrow$ Se) in MnSc$_2$$X_4$ reduces the energy window of excitations for MnSc$_2$Se$_4$ (Fig. \ref{fig:focus_results} (b)).

 \section{Summary and Discussion}{\label{Sec7}}
To summarize, using neutron scattering and macroscopic magnetic magnetic measurements, we unravel complex magnetism of MnSc$_2$Se$_4$ compound. The macroscopic magnetization data infer multiple long-range order transitions at temperatures, which are nearly by one order of magnitude smaller than the Curie-Weiss temperature of 18.4 K indicating the significant frustration. Our neutron powder diffraction studies identify the long-range order as associated with the propagation vector $k$=(3/4 3/4 0).\\
In addition to the presence of long-range order, we found short-range correlations manifested as a diffuse scattering signal that persists towards the high temperatures. In the temperature range of 1.8 K $<T<$ 2 K these short-range correlations co-exist with the long-range ordered state. These are expected to originate from the spiral surface, as predicted for the highly frustrated spiral spin liquid state \cite{bergman_2007}. Thus, MnSc$_2$Se$_4$ undergoes multiple transitions from a high T uncorrelated state for $T\geq |\Theta_{\rm CW}|$, via the spin liquid state for the temperatures between, $T_{N}\leq T\leq |\Theta_{\rm CW}|$, to magnetically long-range ordered states for $T\leq T_{\rm N}$. \\
Our inelastic neutron scattering results fitted to the $J_1$-$J_2$-$J_3$ model, allowed to determine the exchange parameters for MnSc$_2$Se$_4$. The obtained ratio of $J_2$/$J_1$ suggests that the MnSc$_2$Se$_4$ compound is another example of the $A$-site spinel, where the spiral spin liquid state is realized.
Additionally, we suggest that a deeper insight on the magnetic anisotropy and the $H-T$ phase diagram can be obtained with the single crystalline samples. 
Such exploration can be helpful to understand the topological non-trival skrymions discovered in such systems. Our present study unravels a potential candidate from the family of the $A$-site spinels to host these states. \\ 
\begin{table}[h]
\caption{The experimental and theoretical information about the magnetic properties of the MnSc$_2X_4$ ($X$=S, Se) compounds.}
\label{tab_expinfo}
\centering
\small
\begin{tabular}{ccc}
\toprule
 & {MnSc$_2$S$_4$} & {MnSc$_2$Se$_4$}\\
\hline
\hline
$\Theta_{\rm CW}$ & -22.9 K &-18.4 K \\
$f$ & 10& 9\\
$J_1$ & -0.31 K & -0.29 K \\
$J_2$ & 0.46 K & 0.36 K\\
$J_3$ & 0.087 K & 0.047 K\\
$J_2$/$J_1$ & 1.48& 1.24\\
\toprule
\end{tabular}
\end{table}

\begin{acknowledgments}
This work was performed at SINQ, Paul Scherrer Institute, Villigen, Switzerland with a financial support of the Swiss National Science Foundation (Grant No. 200020-182536 and and 206021-139082). This work was partly supported by the Deutsche Forschungsgemeinschaft (DFG) through Transregional Research Collaboration TRR 80 (Augsburg, Munich, and Stuttgart), and by the project ANCD 20.80009.5007.19 (Moldova). We thank S. Gao and S. Nikitin for their help in the analysis of the neutron spectra with spinW and S. Ghara and D. S. Negi for the useful discussions.
\end{acknowledgments}

\end{document}